# Resonant guided wave networks

Eyal Feigenbaum[*] and Harry A. Atwater

*Applied Physics, California Institute of Technology, Pasadena, CA 91125,*

*\* eyalf@caltech.edu*

## Abstract

A resonant guided wave network (RGWN) is an approach to optical materials design in which power propagation in guided wave circuits enables material dispersion. The RGWN design, which consists of power-splitting elements arranged at the nodes of a waveguide network, results in wave dispersion which depends on network layout due to localized resonances at several length scales in the network. These structures exhibit both localized resonances with Q ~ 80 at 1550 nm wavelength as well as photonic bands and band-gaps in large periodic networks at infrared wavelengths.

In the last two decades, several photonic design approaches have defined new directions for control of optical dispersion. Photonic crystals [1-4] exploit periodic structures to create dispersive Bloch wave modes in materials. Metamaterials [5,6] capitalize on the 'meta-atom' concept in which superatomic but subwavelength resonant structures enable complex refractive indices not found in nature. In this Letter, we introduce a synthetic approach to optical dispersion control based on resonant guided wave networks (RGWNs), in which power-splitting elements are arranged in a two-dimensional (2D) or three-dimensional (3D) waveguide network. In a typical RGWN, the photonic elements are designed so that the power is split equally among the waveguides connected to each element, so that a primitive or non-primitive unit of connected waveguides in the network acts as a resonator. Thus RGWNs have features in common with metamaterials and photonic crystals, but also features distinct from each: i) for photonic crystals dispersion stems from nonlocal Bloch wave interference, whereas dispersion in RGWNs arises from local interference of guided waves at network nodes, and ii) unlike metamaterials RGWNs exhibit network dispersion that depends on phase retardation effects *between* rather than *within* elements.

We first illustrate the RGWN concept with a 2D network composed of intersecting metal-insulator-metal (MIM) waveguides, and illustrate how the guided wave network dispersion depends on the waveguide modal properties, waveguide length and network topology. The equal power splitting features of RGWNs are illustrated by the Au-air network in Fig. 1a. MIM waveguides support a lowest-order plasmonic mode in the visible and the near infra-red wavelength range [7-9]. This lowest-order transverse magnetic mode ($TM_0$) does not exhibit modal cutoff, and hence the MIM waveguide supports subwavelength modal cross-sections [10-

12]. It was recently reported that a cross-junction which consists of two normally intersecting MIM waveguides with sub-wavelength gap sizes splits the incoming pulse equally four ways [13]. This equal optical power splitting was observed for continuous waves and also for very short pulses of few optical cycles in the near infrared wavelength range, conserving the pulse shape. In such a waveguide junction, the input waveguide acts as a subwavelength aperture exciting output waveguides in the junction, resulting in a broad spectrum of plane waves with a significant fraction of sideways-going optical power coupled into the waveguides perpendicular to the input waveguide. This nano-aperture effect, which facilitates the equal four-way splitting of optical power, is enabled for transmission lines (e.g., MIM and coaxial configurations) but is mostly limited for purely dielectric waveguides by their half-wavelength modal cross-sections.

A prototypical 2D-RGWN topology consisting of insulating gaps in a metallic bulk constitutes a network of coupled crossed-waveguide junction elements, which we term as "X-junctions". Each X-junction element has four waveguide terminals, serving both as inputs and outputs: when an incoming wave enters through one of them it is split evenly between them and channeled to the four closest neighboring X-junction elements by MIM waveguides. This strong coupling to all four neighboring X-junctions gives this material an optical response different from a cross-coupled network of wavelength-scale purely dielectric waveguides, where most of the power would be transmitted in one direction, with weak coupling between perpendicular waveguides. The rectangular metal cladding regions between the insulating waveguides are much thicker than the skin depth, preventing optical power flow through the material except through the waveguide network. This configuration forms a RGWN using relatively simple elements and network topology. The dimensions of the elements are inherently subwavelength to facilitate four-way equal optical power splitting, and the waveguide lengths are on the scale of the wavelength – resulting in an artificial photonic material on the micro-scale for visible and near infrared wavelengths.

The equal power splitting in X-junction for small gaps was verified by numerical solutions and it was also found that the reflected pulse is out-of-phase (i.e., $\pi$-phase shifted) with the sideways and forward transmitted pulses. The power splitting in the Au-Air X-junction was investigated via 2D finite-difference time domain (FDTD) numerical solutions of Maxwell equations with a short pulse excitation using a central wavelength of 1.5μm, where the Au complex permittivity is fitted to measured data [14]. The gap sizes of the two intersecting MIM waveguides constituting the X-junction are equal. Equal optical power splitting occurs for a MIM insulator gap size below 0.2 μm, as seen in Fig. 1b. As the gap size is increased, the optical power flow deviates from the nano-aperture equal power splitting limit towards the wavelength-scale photonic mode limit, where most of the power is transmitted directly across the X-junction [13]. The data of Fig. 1c indicates that the transmitted and reflected pulses are out-of-phase ($\Phi_R$-$\Phi_F$~$\pi$) and that the phase shift between the sideways-going and the forward transmitted pulses is consistent with the geometrical difference in their pulse propagation trajectories ($r_S$- $r_F$ = ($\Phi_S$-$\Phi_F$)/$k_0 n_{eff}$ = *($\sqrt{2}$-1)d*).

After characterizing the properties of RGWN building blocks, we investigated the dynamics of a small network and show that it forms a resonator. A square (2×2) RGWN is comprised of four X-junctions arranged in a square array, as illustrated in Fig. 2. In order to form a resonator, the X-junction design should enable network dynamics such that pairs of pulses excite the X-junctions out-of-phase, effectively forming mirrors for pulses incident out of phase upon the X-junctions, as illustrated in Fig 2a.

A simplified analytical model shows that in order to maximize the resonant Q for an externally excited (2×2) network, the MIM gap size should not be infinitely small. When the (2×2) RGWN in Fig. 2c is excited from the lower-left arm, the first power splitting event occurs in junction 1, and the second power splitting events occur simultaneously in junctions 2 and 4. The third power splitting event occurs as the pulses arrive at junctions 1 and 3, from junctions 2 and 4. Each junction 1 and 3 is excited simultaneously from two waveguides. The incoming pulses arrive at both junctions in-phase and thus result in destructive pulse interference at junctions 1 and 3. After this power splitting event, the two pulses arriving simultaneously at each junction are out-of-phase and their interference gives rise to strong pulse reflection. Thus the formation of such an effective "mirror" implies that if the pulse has not vanished by the time the third power splitting event, it will enter a resonant state. This illustrates that the effective mirror reflectivity depends on the history of pulse dynamics and the local properties of the X-junction. As the gap size is increased, junction power splitting behavior deviates from the "nano-aperture" limit, and the reflectivity of the effective mirrors is slightly reduced. The trade-off between maintaining balanced power splitting in X-junctions and also high reflectivity of out-of-phase pulses suggests that MIM gap sizes that are subwavelength, but not arbitrarily small, will maximize the network resonance.

The model predictions are verified by 2D electromagnetic simulations, showing that the network dynamics are characterized by an initial transient propagation phase, followed by the development of a resonant state. In Fig. 2c, the simulation snapshots for the first power splitting events are depicted. After a transient which includes the five first splitting events, the resonant state is approached as pairs of pulses resonate between the effective mirrors at junctions 1 and 3 (exemplified by the snapshot $t_6$) and junctions 2 and 4 (exemplified by the snapshot $t_7$). In the third power splitting event, the two pairs of pulses excite the junctions almost in phase, and so only a small fraction of the initial power is coupled into the (2×2) network comprising junctions 1-4. During the fourth and fifth splitting events, there is a strong asymmetry in pulse transmission between the two exciting pulses arriving at each junction, as the power reflected back towards junction 2 is negligible with respect to that reflected towards junction 4.

For the same 2D network topology, but with 3D high aspect ratio Au-air channel plasmon waveguides, the observed wave dynamics closely resembles that for the 2D MIM waveguide networks. The (2×2) 2D network of high aspect ratio channel plasmon waveguides, consisting of rectangular air-core channels in an Au film, was studied with 3D full field simulations (Fig. 2d). If the aspect ratio of the channel plasmon waveguide is high enough, the propagating mode within the channels strongly resembles the MIM gap plasmonic mode [15,16]. The measured resonant network $Q = 82$ at a wavelength of 1.5 μm obtained from 3D simulations for channel plasmon waveguides is in a good agreement with the value $Q = 83$ obtained from 2D simulations (MIM waveguides). Snapshots in Figs. 2d illustrate the two power splitting events that define the resonant state for the channel waveguides, closely resemble those seen in Fig. 2c for the MIM waveguides, and also corresponding to the ones in the principal illustration in Fig. 2b.

The Q-factor values of the RGWN in Fig. 3a illustrate the role of the effective mirrors, for which the power can be completely conserved, in generating strong network resonances. The network Q-factor obtained is an order of magnitude larger than what would be expected if optical power splitting in the X-junctions operated incoherently, i.e., disregarding interference effects and thus losing half of the power in each splitting event. We also see that the network Q-factor decreases with increasing the gap size, resulting in a reduction in the reflectivity of the effective mirrors. On the other hand, as the gap size is decreased, plasmonic mode attenuation increases due to metallic waveguide losses, and so a maximal Q-factor value is obtained for a gap size of 250nm. Those Q-factor values are comparable to the ones (Q~100) only due to radiation loss for the case of cylindrical dielectric cavity with refractive index of n = 2.5 and k = 0 surrounded with air [17], and with similar dimensions in terms of cavity wavelength to the (2×2) RGWN discussed above (radius=1.3λ). If we were to artificially decrease the Au loss at 1.5 μm (or alternatively at higher wavelengths), the Q-factor of the resonator would increase tremendously (e.g., Q = 750 for 200nm gap width), indicating that the resonator Q factor is limited by the material loss at 1.5 μm.

RGWNs have features that are also broadly reminiscent of Mach-Zehnder interferometers. In Fig. 3b the effect on the Q-factor of introducing a 1 μm long dielectric region into the gap of one of the waveguides is presented. It indicates that while the channel plasmon RGWN may be sensitive to dielectric gap defects and inhomogeneities, the Q-factor does not drop abruptly to zero in the presence of defects, which exemplifies the network robustness to possible fabrication artifacts.

After studying the resonant network physics in a small (2×2) RGWN, we further investigated the dispersion characteristics infinite 2D periodic waveguide networks. We find that RGWNs exhibit wave dispersion and photonic bandgaps due to interference effects, and that the band-structure can be controlled by modifying the network structural parameters. Two different length-scales control the dispersion properties of the network: the phase shift properties of X-junctions are set by the width of the MIM gaps, and the interference scheme is determined by the distance between nodes and their arrangement according to the network topology. If the network parameters are chosen such that plane wave excitation at a given incidence angle results in a resonance effect similar to the one demonstrated for the (2×2) network, this would correspond to a forbidden state for propagation within the photonic bandgap on the dispersion diagram. Examining the optical density of states (DOS) for different wave vectors over frequencies in the near infrared range, where the material dispersion is small, we observe a photonic band structure which is only due to dispersion resulting from the network topology, as shown in Fig. 4a. The photonic band structure depends directly on the inter-node distance (i.e., the waveguide length) as seen in Fig. 4b, which determines the wave retardation between junctions, and therefore is scalable with network dimensions. The MIM gap size does not exhibit the same dependence; instead the band energy varies inversely with MIM waveguide gap size. Further possibilities for achieving band dispersion control are illustrated in Fig. 4c, where we observe the appearance of flat bands over wide range of k-vectors at 130 THz and 170 THz, as well as the formation of a photonic bandgap between 140-160 THz, for appropriately chosen network parameters.

The dispersion design in a volume can be addressed by 3D-RGWN topologies, for example, constructing an array of orthogonally intersecting 3D networks of coaxial Au-air waveguides, each aligned with a different Cartesian axis (Fig. 5a). In this case, the four arm X-junction element of the 2D network is replaced by a six arm junction element in a 3D network. Coaxial transmission lines support a $TM_0$ mode, which does not exhibit mode cut-off, enabling equal power splitting between the six arms. We have verified that six-way equal power splitting occurs for pulsed excitation in coaxial Au-air waveguide junctions, using 3D electromagnetic simulations. The band-diagrams for infinitely large periodic 3D-RGWNs of such six-arm junctions illustrate the role of network parameters, such as waveguide length, in determining the optical dispersion (Fig. 5b and 5c), as the waveguide and its dispersion are the same for both cases.

RGWNs represent a new approach for design of dispersive photonic materials. While simple examples have been illustrated here, we also suggest that in the most general case of a RGWN, the power splitting elements and waveguide lengths would not need to be homogenous or isotropic across the network; the dispersion-controlling parameters (i.e., gap size and inter-node distance) could be varied throughout the network in both dimensions. RGWNs are distinctly different from photonic crystals, which rely on the formation of Bloch wave states by interference of waves diffracted from an array of periodic elements, which is truly a non-local phenomenon. By contrast, a RGWN coherently superposes power flowing along isolated waveguides at X-junctions. Furthermore, in a photonic crystal, the interference pattern of the diffracted waves depends on the *nonlocal* periodic spatial arrangement of the diffracting elements, in RGWNs the *local* network topology determines the dispersion and resonance features. For example, in the RGWN, there is no restriction on whether the waveguides are straight or curved; only the total path length along the waveguide and the phase shift upon power splitting determine the coherent wave propagation through the network. Metamaterials also feature a design approach based on the attributes of localized resonances, but the dispersive properties of metamaterials do not depend on any length scale between resonant elements – thus differing substantially from RGWNs. Finally we note the differences between RGWNs and arrays of coupled resonator optical waveguides [18]. Coupled resonator optical waveguides feature discrete identifiable resonators that act as the energy storage elements, and dispersion occurs as modes of adjacent resonators are evanescently coupled. By contrast, in RGWNs, energy is not stored resonantly in discrete resonators, but in the network of waveguides that are designed to exhibit resonant behavior as a network. While previously reported resonant plasmonics circuits (e.g., [19-22]) have fundamentally one-dimensional topologies (input-device-output), the RGWNs are fundamentally different being 2D and 3D constructs and thus mediate and regulate power flow in a *material* rather than in a discrete device.

## Figure captions

**Figure 1:** (a) 2D RGWN implementation with an array of crossed MIM waveguides and two possible resonant loop trajectories (dashed lines). Power splitting properties of the emerging pulses in an X-junction: (b) intensity relative to the exciting pulse, and (c) phase difference. Pulsed excitation is at a central wavelength of $\lambda_0=1.5\mu m$ with $0.36\mu m$ band-width (26fs).

**Figure 2:** Resonance buildup in a (2×2) RGWN. (a) Two out of phase input pulses result in two output pulses into the same waveguides from which the pulses entered the junction, and the X-junction acts effectively as a mirror (+/- correspond to two $\pi$ shifted phase states of pulse). (b) steady states of waves resonating in a (2×2) network where each pair of pulses excites the X-junctions out of phase, and in the absence of loss, waves resonate indefinitely in the network. MIM RGWN: (c) Schematics and snapshots of $H_z$ (normalized to the instantaneous maximal value) at the third to the seventh power splitting events for a 2D-FDTD simulation. $d=0.25\mu m$, $L=6\mu m$. Channels RGWN: (d) 3D-FDTD of 2D topography of air channel waveguides network is excited by $E_x$-polarized; snapshots of the H-field in the parallel plane $0.1\mu m$ underneath the Au-air interface ($z=0.9\mu m$) and in the plane normal to the Au-air interface (y=0). The resonant state for different 2D-RGWN implementations are evident in (b), $t_6$ and $t_7$ in (c), and in-plane panel in (d). $\lambda_0=1.5\mu m$.

**Figure 3:** Q-factors of Au/air MIM (2×2) RGWN resonator. (a) Q-factors from simulation results compared with those resulting from incoherent power splitting. The Q-

factor contribution of the material loss is calculated from the attenuation coefficient [17,23]. (b) network robustness with respect to dielectric defects: Q-factor vs. the refractive index of a 1μm long defect. d=0.25, L=4μm, $\lambda_0$=1.5μm.

**Figure 4:** Photonic band structure of an infinitely large periodic RGWN: (a) optical DOS for a square periodic unit cell with d=0.25μm, L=3μm. (b), dependence of the central wavelength of the photonic band (at Γ point at ~180THz in Fig. a) on L for d=0.3μm, and on g for L=4μm. (c), Photonic band gap formed in the optical DOS in rectangular periodic unit cell. In the x-direction: d=0.24μm, L=2μm; in the y-direction: d=0.2μm, L=1.2μm.

**Figure 5:** 3D RGWN: (a) rendering of a 3D RGWN building block (6-arm junction). Optical DOS of an infinite 3D network spaced periodically with cubic periodic unit cell (b) L=2μm and (c) L=2.2μm.

# Figure 1

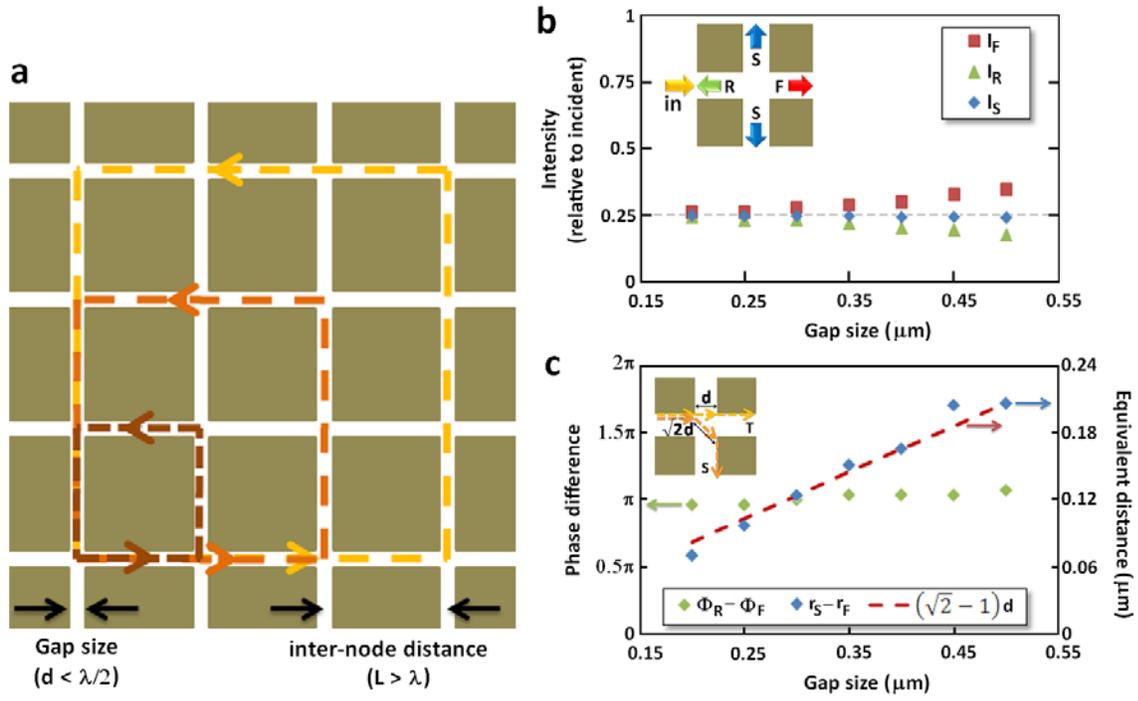

# Figure2

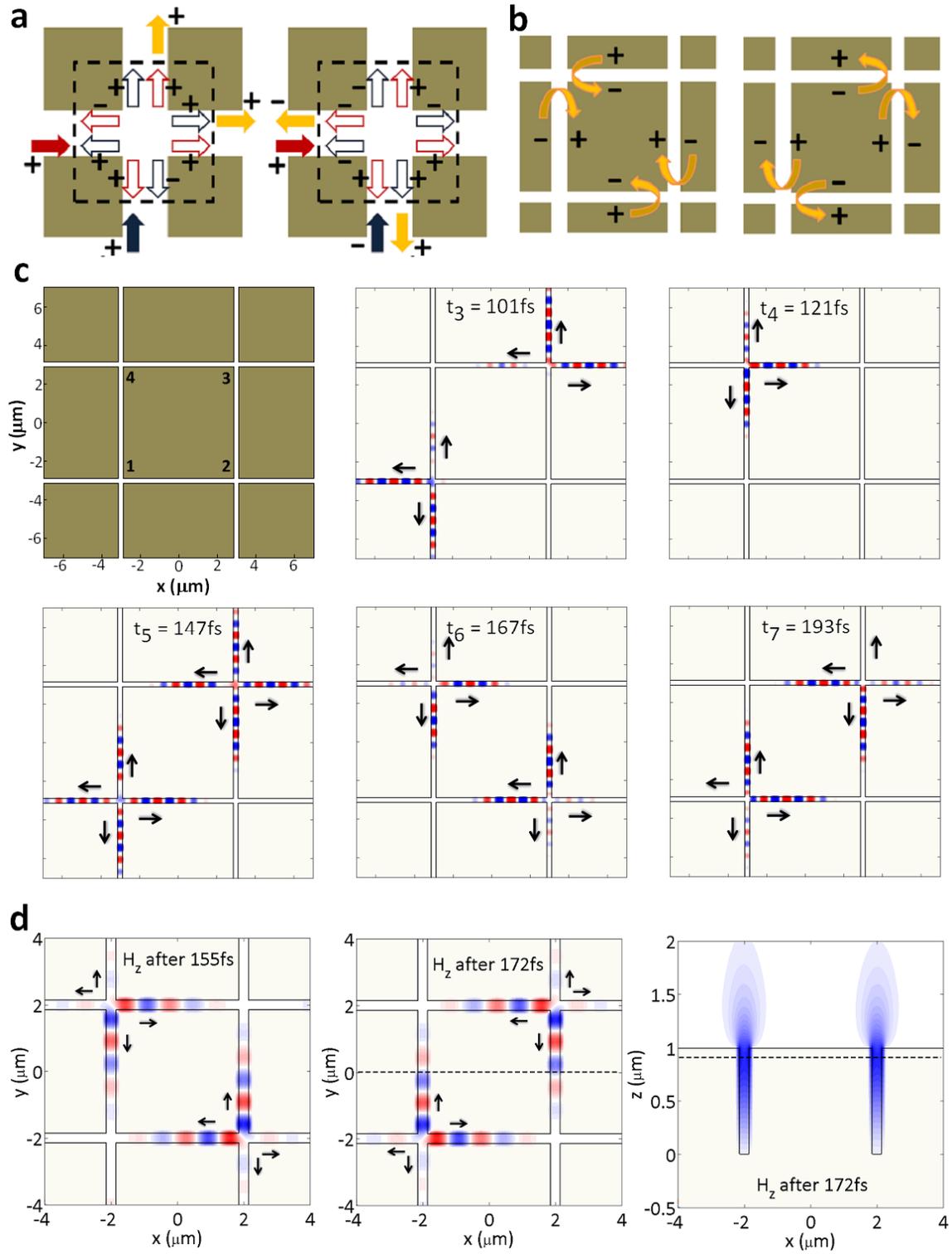

# Figure 3

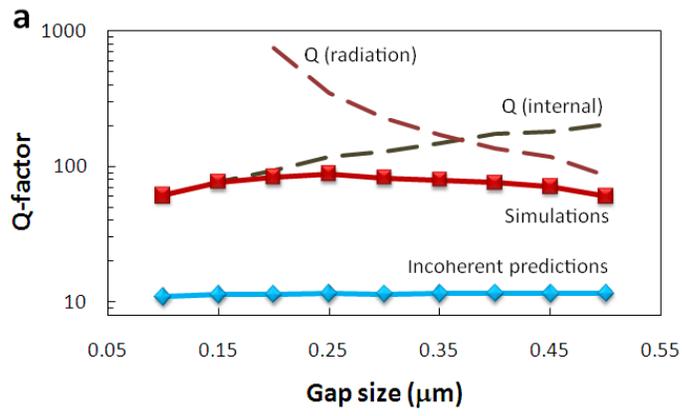 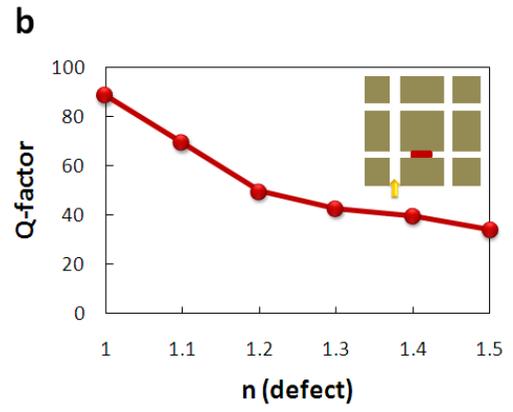

# Figure 4

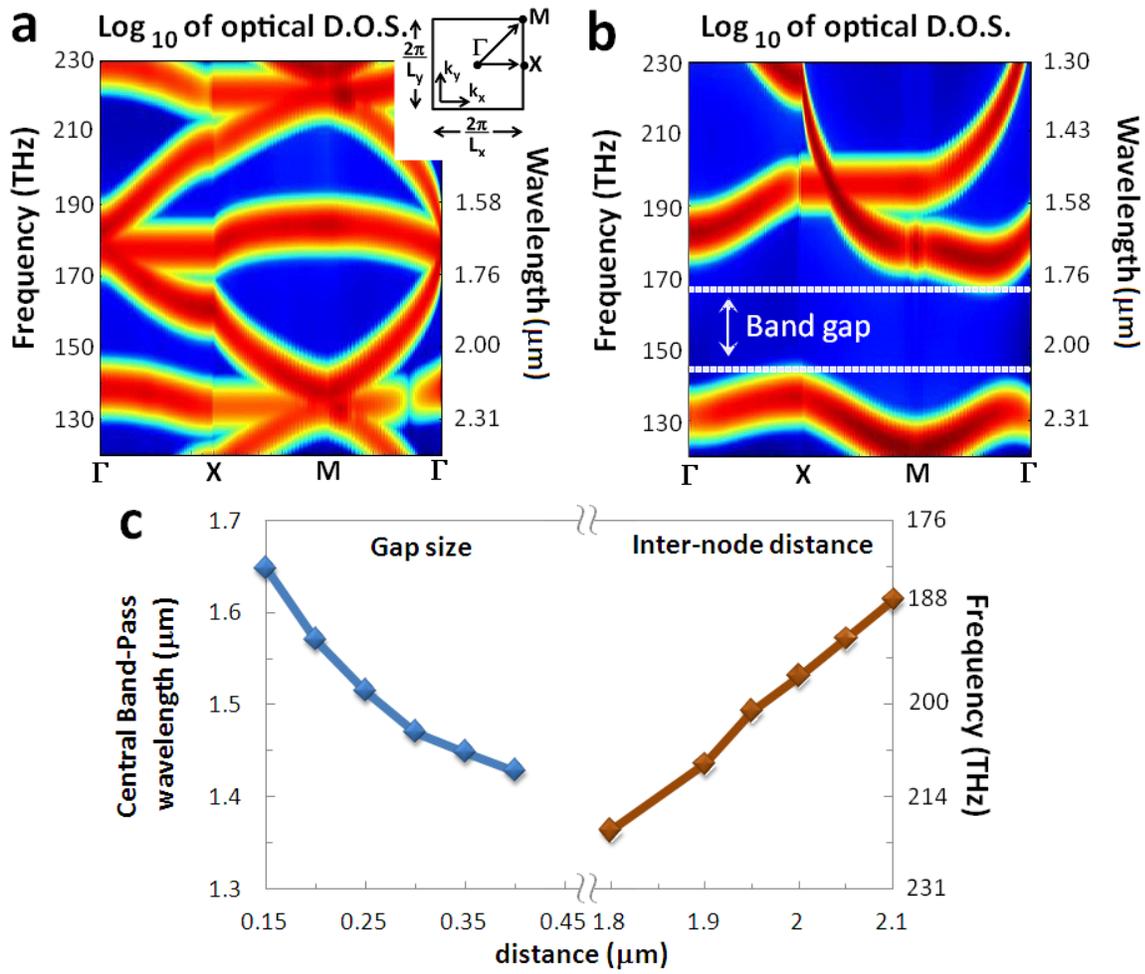

# Figure 5

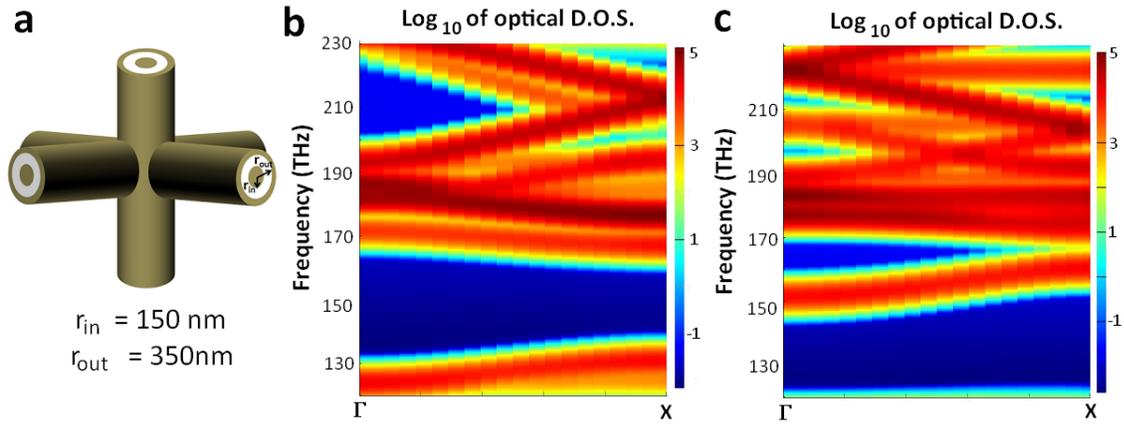

$r_{in}$ = 150 nm
$r_{out}$ = 350nm